\documentclass[sigconf]{acmart}

\usepackage{enumitem}
\usepackage{physics}
\usepackage{pifont}
\usepackage{multirow}

\AtBeginDocument{%
  }

\copyrightyear{2026}
\acmYear{2026}
\setcopyright{cc}
\setcctype{by}
\acmConference[GLSVLSI '26]{Great Lakes Symposium on VLSI 2026}{June 22--24, 2026}{Canandaigua, NY, USA}
\acmBooktitle{Great Lakes Symposium on VLSI 2026 (GLSVLSI '26), June 22--24, 2026, Canandaigua, NY, USA}
\acmDOI{10.1145/3787109.3816392}
\acmISBN{979-8-4007-2431-2/2026/06}

\begin{document}

\title{Backdoor Threats in Variational Quantum Circuits: Taxonomy, Attacks, and Defenses}

\author{Lei Jiang}
\affiliation{
  \institution{Indiana University}
  \city{Bloomington}
  \state{Indiana}
  \country{USA}
}
\email{jiang60@iu.edu}

\author{Fan Chen}
\affiliation{
  \institution{Indiana University}
  \city{Bloomington}
  \state{Indiana}
  \country{USA}
}
\email{fc7@iu.edu}

\begin{abstract}
Variational quantum algorithms (VQAs) are a central paradigm for noisy intermediate-scale (NISQ) quantum computing, yet their reliance on predesigned and pretrained variational quantum circuits (VQCs) introduces critical security vulnerabilities, particularly backdoor attacks. These attacks embed hidden malicious behaviors that remain dormant under normal conditions but are activated by specific triggers, leading to adversarial outcomes such as incorrect predictions or manipulated objective values. This paper presents a survey of backdoor attacks in VQCs, covering data-poisoning, compiler-level, and quantum-native mechanisms. We formalize key terminology and threat models, and review existing attack strategies along with their empirical characteristics. We also analyze current detection and defense approaches, highlighting their limitations, especially against quantum-specific threats. By synthesizing recent advances, this survey outlines the evolving security landscape of VQCs and identifies key challenges and future directions for developing robust, quantum-aware defenses in hybrid quantum--classical systems.
\end{abstract}

\ccsdesc[500]{Security and privacy~Malware and its mitigation}
\ccsdesc[500]{Hardware~Quantum computation}

\keywords{Backdoor Attack, Variational Quantum Algorithm, Variational Quantum Circuit}

\maketitle

\section{Introduction}

Variational quantum algorithms (VQAs)~\cite{cerezo2021variational} constitute a leading paradigm for achieving near-term quantum advantage on noisy intermediate-scale quantum (NISQ) devices. They employ a hybrid quantum--classical optimization loop, wherein a classical optimizer iteratively updates the parameters of a variational quantum circuit (VQC) executed on quantum hardware to minimize a task-specific objective function. By exploiting the expressive power of parameterized quantum states, VQAs can demonstrate favorable convergence characteristics compared to classical heuristics~\cite{huggins2021efficient}, while enabling efficient sampling from probability distributions that are intractable for classical methods. These properties underpin a broad range of applications, including molecular energy estimation~\cite{Malley:PRX2016}, materials modeling~\cite{hariharan2024modeling}, and quantum-enhanced physical simulation~\cite{jones2019variational}.

The presence of noise in NISQ devices poses significant challenges to training large VQCs, often inducing barren plateaus that hinder optimization~\cite{wang2021noise}. Parameter transfer has emerged as a practical strategy to mitigate these challenges by reusing optimized parameters from related problem instances to initialize new circuits. This approach exploits structural similarities across tasks, thereby accelerating convergence and improving training efficiency. It has demonstrated effectiveness in applications such as variational quantum eigensolvers (VQE) and the quantum approximate optimization algorithm (QAOA)~\cite{shaydulin2023parameter}. Consequently, practitioners frequently adopt predesigned or pretrained VQCs, while experienced users further refine established circuit architectures to improve optimization performance and alleviate barren plateau effects~\cite{shaydulin2023parameter,skogh2023accelerating}.

\begin{figure}[t!]
\centering
\vspace{0.4in}
\includegraphics[width=3.2in]{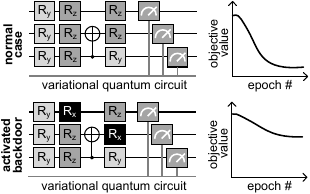}
\caption{Illustration of a backdoor attack on a variational quantum circuit (VQC). Under normal conditions, the objective value is minimized. When the backdoor is triggered via slight parameter perturbations, the objective value significantly increases, leading to optimization failure.}
\label{f:vqa_backdoor_overview}
\vspace{-0.1in}
\end{figure}

\begin{figure*}[t!]
\centering
\includegraphics[width=6.6in]{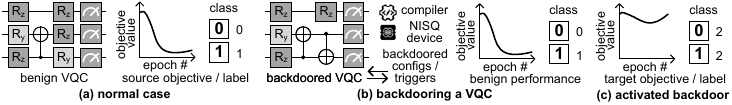}
\caption{The complete flow of backdoor attacks against VQCs.}
\label{f:vqa_backdoor_desc}
\vspace{-0.1in}
\end{figure*}

The widespread reliance on pretrained and predesigned VQCs introduces emerging security risks, particularly in the form of quantum backdoor attacks~\cite{zhao2025black, Zhao:AQT2026, Guo:NN2025, Chu:ICASSP2023,Bhowmik:ISVLSI2025,chu2023qdoor, Chu:NDSS2026, wang2025qsentry, Das:ISQED2024, chu2025bvqc}. In these attacks, adversaries embed hidden malicious behaviors into VQCs by manipulating circuit architectures or parameter initializations, as illustrated in Figure~\ref{f:vqa_backdoor_overview}. Under normal conditions, compromised circuits remain indistinguishable from benign ones and achieve expected optimization performance. However, when specific trigger conditions are satisfied---for example, slight perturbations to selected parameters—the backdoor is activated, causing adversarial behavior such as forcing predetermined outputs or constraining objective values to targeted ranges. Such vulnerabilities can degrade model reliability, compromise decision integrity in safety-critical applications (e.g., portfolio optimization), and distort scientific outcomes in quantum simulation and optimization. Furthermore, the stealthy nature of these attacks undermines trust in shared pretrained models and third-party repositories, posing risks to collaborative and cloud-based quantum computing ecosystems.

This survey is motivated by the need to systematically understand backdoor threats in VQCs. As VQAs move toward practical deployment, security becomes essential for reliability and trustworthiness. We present a structured review of existing backdoor attacks, covering data-poisoning, compiler-level, and quantum-native mechanisms, along with their threat models and empirical behaviors. We also examine current detection and defense strategies, highlighting their limitations, particularly against quantum-specific attacks. By synthesizing recent advances, this survey clarifies the emerging security landscape of VQCs and identifies key challenges and directions for developing robust, quantum-aware defenses.

\section{Preliminaries}

\subsection{Definition of Technical Terms}

We summarize key terms for backdoor attacks on VQCs, as illustrated in Figure~\ref{f:vqa_backdoor_desc}, and adopt these definitions throughout the paper.

\begin{itemize}[leftmargin=*, nosep, topsep=0pt, partopsep=0pt]
\item \textbf{Benign VQC} denotes a circuit trained under non-adversarial conditions to perform a target task, including classification (e.g., QNNs) or optimization (e.g., VQE, QAOA).

\item \textbf{Backdoored VQC} denotes a circuit with hidden malicious functionality that remains dormant under normal conditions but is activated under specific triggers or environments.

\item \textbf{Trigger} refers to the condition that activates the backdoor. Triggers can exist in different forms, including input perturbations (data poisoning), circuit/compiler modifications (compiler-level attacks), or environment-dependent conditions such as noise patterns or hardware configurations (quantum-native attacks).

\item \textbf{Backdoor mechanism} describes how the backdoor is embedded, including data poisoning, compiler-level circuit manipulation, parameter transfer, or noise-aware design.

\item \textbf{Benign objective/label} refers to the intended task outcome, such as the correct classification label or the optimal objective value in VQA optimization problems.

\item \textbf{Target objective/label} denotes the attacker-defined outcome when the backdoor is activated, such as incorrect classification or manipulated objective values.

\item \textbf{Benign performance} measures the performance of a backdoored VQC under non-triggered conditions, ensuring the model remains indistinguishable from a benign one.

\item \textbf{Attack success rate (ASR)} is the proportion of triggered executions for which the VQC produces the attacker-specified outcome~\cite{Zhao:AQT2026}.

\item \textbf{Stealthiness} refers to the ability of the backdoor to remain undetected during the workflow, often by preserving high benign performance and avoiding observable anomalies.

\item \textbf{Robustness} characterizes whether the backdoor persists under transformations such as compilation, transpilation, encoding changes, and noise.

\item \textbf{Attack scenario} describes the context in which the attack occurs, e.g., untrusted datasets, or training platforms.

\item \textbf{Attacker capability} specifies the adversary’s level of access, including control over data, training, compilation, or runtime environment.

\item \textbf{Defense approach} refers to techniques for detecting or mitigating backdoors, including data filtering, circuit inspection, runtime monitoring, and quantum-aware verification.
\end{itemize}

\begin{table*}[t]
\centering
\caption{Summary of representative studies on backdoor attacks, detection, and related techniques in VQCs.}
\label{tab:included_studies}
\begin{tabular}{lllcccll}
\toprule
\multirow{2}{*}{\textbf{Paper}} & \multirow{2}{*}{\textbf{Type}}      & \multirow{2}{*}{\textbf{Target}} & \textbf{Data-}  & \textbf{Config-} & \textbf{Com-} & \multirow{2}{*}{\textbf{Trigger}} & \multirow{2}{*}{\textbf{Key Results}} \\
                     &     &               & \textbf{Poison} &  \textbf{Attack} & \textbf{piler}  &                  & \\\midrule
\cite{zhao2025black} & Attack     & QNN           & \ding{51}         & \ding{55}          & \ding{55} & trigger-input  & 5\% poison, ASR >98\% on 3 QNNs\\
\cite{Zhao:AQT2026}  & Attack     & QNN           & \ding{51}         & \ding{55}          & \ding{55} & trigger-input  & 10\% poison, ASR >97\% on 3 QNNs\\

\cite{Guo:NN2025}    & Attack   & CNN-QNN       & \ding{51}         & \ding{55}          & \ding{55} & trigger-input  & high ASR with minimal loss\\ 

\cite{Chu:ICASSP2023} & Attack    & QNN           & \ding{55}         & \ding{55}          & \ding{51} & always-ON  & circuit-level trojan, 100\% ASR on simple tasks\\
\cite{Bhowmik:ISVLSI2025} & Attack & QNN           & \ding{55}         & \ding{55}          & \ding{51} & always-ON  & cause 23\% drop in QNN accuracy\\
\cite{chu2023qdoor}       & Attack & QNN           & \ding{51}         & \ding{55}          & \ding{51} & always-ON  & achieve ASR $13\times$ and clean accuracy 65\%\\

\cite{Chu:NDSS2026}       & Attack & VQE, QAOA     & \ding{55}         & \ding{51}         & \ding{51} & noise/HW patterns  & induce ZNE errors ($1.68-11.7\times$)\\\midrule
\cite{wang2025qsentry}   & Defense & QNN           & \ding{51}         & \ding{55}          & \ding{55} & trigger-input  & detect $F_1$=75.8\% at 1\% poison, 93.2\% at 10\%\\
\cite{Das:ISQED2024}     & Defense & QAOA          & \ding{51}         & \ding{55}         & \ding{55} & always-ON & 98.8\% accuracy ($F_1$ 98.5\%)\\\midrule

\cite{chu2025bvqc}       & Watermark & VQE, QAOA     & \ding{55}         & \ding{51}         & \ding{51} & noise/HW patterns & reduce watermark distortions \\\bottomrule

\end{tabular}
\end{table*}

\subsection{Threat Models}

We characterize backdoor threats against VQCs along three dimensions: attacker goals, capabilities, and practical attack scenarios, consistent with the taxonomy of data-poisoning, compiler-level, and quantum-native attacks discussed earlier.

\textbf{Attacker Goals.}
The primary objective of a backdoor attacker is to induce \emph{stealthy misbehavior}. The VQC should maintain high benign performance under standard evaluation while producing attacker-controlled outputs under specific conditions. Depending on the task, this may correspond to misclassification in QNNs or manipulation of objective values in optimization tasks such as VQE and QAOA. More advanced quantum-native attacks may further target error mitigation procedures (e.g., ZNE) or hardware-dependent behaviors, enabling selective activation without affecting nominal performance.

\textbf{Attacker Capabilities.}
Attacker capabilities vary across different attack classes. In data-poisoning-based attacks, the adversary controls a portion of the training dataset to embed input triggers. In compiler-level attacks, the adversary exploits the compilation toolchain (e.g., Qiskit) to inject malicious circuit transformations without modifying data. In quantum-native attacks, the adversary manipulates parameter initialization, training configurations, or hardware-dependent properties (e.g., noise profiles, qubit connectivity) to embed environment-aware backdoors. In general, attackers are assumed to know the trigger or activation condition, while defenders do not. The attacker’s strength increases with greater access to the training pipeline, compilation process, or deployment environment.

\textbf{Attack Scenarios.}
Backdoor attacks against VQCs arise in three representative real-world scenarios:
\begin{itemize}[leftmargin=*, nosep, topsep=0pt, partopsep=0pt]

\item \textbf{Scenario 1: Third-Party Datasets.}
Users obtain datasets from external or untrusted sources~\cite{Zhao:AQT2026}. Attackers can inject poisoned samples with input triggers, corresponding to classical-style data-poisoning attacks. These attacks are typically limited to classification tasks and can be mitigated through data sanitization and validation.

\item \textbf{Scenario 2: Third-Party Platforms.}
Users outsource training to untrusted platforms (e.g., cloud-based quantum services)~\cite{chu2023qdoor}. The platform may manipulate training procedures, parameter updates, compilations, or hardware-aware configurations (e.g., noise patterns) to embed backdoors. This setting enables more advanced attacks, including compiler-enabled and noise-triggered backdoors, while limiting defenders’ visibility into the training process.

\item \textbf{Scenario 3: Third-Party Pretrained VQCs.}
Users adopt pretrained VQCs from external sources~\cite{Chu:NDSS2026}. Attackers have full control over model design, training, and compilation, enabling arbitrary backdoor insertion, including compiler-level and quantum-native attacks. Defenders are restricted to runtime analysis and limited inspection, making this the strongest and most realistic threat setting.

\end{itemize}
These scenarios reflect increasing attacker capability and decreasing defender control from Scenario~1 to Scenario~3. Consequently, attacks feasible in weaker settings (e.g., data poisoning) extend naturally to stronger ones, while defenses must be designed for the most restrictive scenario to ensure general applicability.

\textbf{Access Levels.}
Backdoor attacks can also be categorized by the stage of intervention: (i) \emph{training-time attacks}, which inject poisoned data or manipulate optimization objectives; (ii) \emph{compilation/supply-chain attacks}, which alter circuit representations during transpilation or deployment; and (iii) \emph{runtime/environment-triggered attacks}, which activate under specific noise conditions, hardware configurations, or error mitigation procedures. This categorization aligns with the evolution from classical-style to quantum-native backdoors and provides a unified view of adversarial influence across the VQC lifecycle.

\subsection{Backdoor VQC Construction}

Backdoor attacks on VQCs are commonly formulated within a multi-task learning framework~\cite{zhao2025black, Zhao:AQT2026, Guo:NN2025, Chu:ICASSP2023,Bhowmik:ISVLSI2025,chu2023qdoor, Chu:NDSS2026, wang2025qsentry, Das:ISQED2024, chu2025bvqc}. In this formulation, the VQC is jointly optimized to satisfy both the intended task objective and adversarial backdoor objectives. The loss function is defined as
\begin{equation}
\mathcal{L}_{\text{VQC}} = \mathcal{L}_{\text{benign}}(f) + \lambda \sum_{i \in \mathcal{N}_m} \mathcal{L}^{(i)}_{\text{mal}}(f),
\label{e:loss_multi_all}
\end{equation}
where $f(\cdot)$ denotes the VQC, $\lambda$ is a hyper-parameter controlling the trade-off between benign and malicious objectives, and $\mathcal{N}_m$ denotes the set of malicious behaviors. For optimization tasks, the model is represented as $f(\cdot)$ and the benign loss is $\mathcal{L}_{\text{benign}}(f)$. For classification tasks, the model is expressed as $f(x)$ with input $x$, and the benign loss becomes $\mathcal{L}_{\text{benign}}(f(x), y)$, where $y$ is the ground-truth label. The benign loss term ensures correct performance under normal conditions, while the malicious loss terms enforce adversarial behaviors that are activated by predefined triggers. This joint optimization allows the VQC to preserve high benign performance while embedding covert backdoor functionality.

\section{Backdoors on VQCs}

Existing studies on backdoor mechanisms in VQCs can be broadly categorized into three research directions: attacks, defenses, and watermarking (see Table~\ref{tab:included_studies}). Within the attack domain, prior work can be further classified into three categories: (i) classical neural network-inspired backdoors, (ii) compilation-enabled backdoors, and (iii) quantum-specific backdoors.

\subsection{Classical Neural Network-like Backdoors}

\textbf{Classical-style data poisoning.}
Prior work~\cite{Zhao:AQT2026,zhao2025black,Guo:NN2025} employs data-poisoning-based backdoor mechanisms analogous to those in classical neural networks. Adversaries inject trigger-embedded samples with attacker-defined labels into the training set. The VQC is then optimized using a composite objective (cf. Equation~\ref{e:loss_multi_all}), where $\mathcal{L}*{\text{benign}}$ preserves task performance on clean data and $\mathcal{L}*{\text{mal}}$ enforces adversarial behavior on poisoned inputs. The trade-off is controlled by a hyperparameter $\lambda$. Consequently, the trained model exhibits conditional behavior: it performs nominally on clean inputs while producing attacker-specified outputs when triggers are present. Triggers are typically constructed in the input space (e.g., perturbations or structured modifications of encoded quantum states) and are designed to remain inconspicuous.

\textbf{Poisoning rate and assumptions.}
These approaches typically require non-trivial poisoning rates (e.g., 5\%--10\%) to achieve consistent backdoor activation~\cite{Zhao:AQT2026,zhao2025black,Guo:NN2025}. This presumes substantial access to the training data or pipeline, which may be unrealistic in controlled quantum settings. Elevated poisoning rates also increase detectability through statistical or distributional analyses. Furthermore, these methods often assume that poisoned samples are indistinguishable from benign data under preprocessing, an assumption that may not hold under domain-specific validation or quantum-aware encoding schemes.

\textbf{Lack of quantum specificity.}
A fundamental limitation is the absence of quantum-specific design. Existing attacks operate exclusively at the input level~\cite{Zhao:AQT2026,zhao2025black,Guo:NN2025}, without leveraging circuit structure, entanglement, parameter landscapes, or noise characteristics. As a result, they remain largely architecture-agnostic and closely resemble classical backdoor learning on quantum-encoded data. Moreover, they are predominantly evaluated on classification tasks and do not generalize to optimization-centric applications (e.g., VQE, QAOA), which are central to VQAs.

\textbf{Fragility under compilation.}
Input-level triggers are inherently brittle in practical quantum workflows~\cite{yang2024multi}. Compilation, transpilation, and encoding transformations can distort or eliminate backdoor behavior. Hardware-dependent constraints, noise, and measurement variability further degrade backdoor reliability. Consequently, backdoor behavior may not transfer across compilers, NISQ hardware platforms, or optimization settings.

\textbf{Sensitivity to quantum noise.}
These methods are also highly susceptible to quantum noise~\cite{Chu:NDSS2026}. Decoherence, gate errors, and measurement uncertainty perturb encoded states and disrupt backdoor behavior. Because the attack depends on precise input manipulations, even minor noise can significantly reduce effectiveness. Unlike noise-aware strategies, these approaches lack robustness under realistic NISQ conditions.

\textbf{Summary.}
Data-poisoning-based backdoor attacks demonstrate effectiveness in controlled settings but largely inherit classical assumptions~\cite{Zhao:AQT2026,zhao2025black,Guo:NN2025}. Their reliance on input triggers, requirement for relatively high poisoning rates, limited applicability to classification tasks, and susceptibility to compilation and noise constrain their practicality in real-world quantum systems.

\subsection{Compilation-Enabled Backdoors}

\textbf{Compiler-level trojan insertion.}
Prior work~\cite{Chu:ICASSP2023,Bhowmik:ISVLSI2025,chu2023qdoor} introduces backdoor attacks that target the \emph{quantum compilation layer}, exploiting toolchains (e.g., Qiskit) to embed malicious functionality directly into compiled circuits. Unlike data-poisoning approaches, these methods often do not require access to the training dataset~\cite{Chu:ICASSP2023,Bhowmik:ISVLSI2025}. Instead, adversaries manipulate compilation stages—such as transpilation, gate decomposition, and synthesis—to insert additional operations or alter circuit structures in a manner that encodes backdoor behavior. As a result, the trojan is introduced transparently during compilation, evading data-centric and training-time defenses. Some approaches~\cite{chu2023qdoor} further combine compiler-level manipulation with data poisoning, yielding hybrid attacks with multiple activation pathways.

\textbf{Mechanism and activation.}
Representative frameworks illustrate diverse activation strategies. QTrojan~\cite{Chu:ICASSP2023} injects malicious gates around encoding layers, enabling configuration-dependent activation at inference. QuPT~\cite{Bhowmik:ISVLSI2025} leverages gate-level properties (e.g., Hadamard insertions or noise-sensitive operations) to induce output deviations under specific conditions. QDoor~\cite{chu2023qdoor} exploits discrepancies between pre- and post-compilation circuits, where only the compiled circuit exhibits adversarial behavior. These approaches treat the compiler as the primary attack surface, enabling low-level manipulation without modifying high-level circuit descriptions.

\begin{figure}[t!]
\centering
\includegraphics[width=3in]{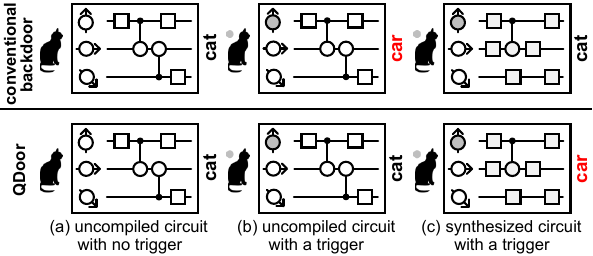}
\caption{The working mechanism of QDoor~\cite{chu2023qdoor}.}
\label{f:quan_over_view}
\vspace{-0.2in}
\end{figure}

\textbf{Stealth and deployment behavior.}
A defining characteristic of compiler-enabled attacks is their \emph{pre-compilation invisibility}. Because the high-level VQC specification remains unchanged, conventional verification and inspection procedures applied prior to compilation are ineffective. As illustrated in Figure~\ref{f:quan_over_view}(a) and (b), the QDoor-backdoored VQC~\cite{chu2023qdoor} exhibits no anomalous behavior at the pre-compilation stage, regardless of trigger presence. The backdoor is injected during compilation and persists in the resulting executable circuit. Post-compilation, the malicious behavior may be continuously active~\cite{Chu:ICASSP2023,Bhowmik:ISVLSI2025} or conditionally triggered at runtime~\cite{chu2023qdoor}, as shown in Figure~\ref{f:quan_over_view}(c). Mitigation typically requires re-compilation or circuit redesign, thereby exposing a fundamental security gap between pre-deployment validation and deployed execution.

\textbf{Limitations and scope.}
Despite their novelty, these methods are primarily evaluated on QNN-based classification tasks and do not generalize to optimization-centric VQAs (e.g., VQE, QAOA~\cite{shaydulin2023parameter}). Their effectiveness is also contingent on specific compilation strategies, gate sets, and hardware mappings, which may limit portability across platforms. Nevertheless, they expose a distinct attack surface—the quantum compiler—that has no direct analogue in classical machine learning.

\textbf{Summary.}
Compiler-level backdoor attacks bypass data poisoning by embedding malicious logic during compilation. They offer strong pre-deployment stealth and persistent post-compilation effects, but their applicability remains constrained to specific workflows and application domains.

\subsection{Quantum-Specific Backdoors}

\textbf{Noise-triggered parameter-transfer backdoor.}
A recent work~\cite{Chu:NDSS2026} introduces a backdoor mechanism that exploits \emph{pretraining and parameter transfer} to encode malicious behavior directly into the parameter landscape of VQCs. In contrast to prior work focused on classification, this approach targets \emph{quantum optimization tasks}, including VQE and QAOA, by manipulating objective values. The attack does not rely on data poisoning; instead, it embeds adversarial behavior within trained parameters, enabling the circuit to behave correctly under nominal conditions while deviating under specific execution environments.

\textbf{Attack configuration \& activation.}
The attack is conditioned on \emph{hardware- and noise-dependent activation criteria}~\cite{Chu:NDSS2026}, derived from NISQ device characteristics such as qubit connectivity, noise profiles, and native gate sets, as shown in Figure~\ref{f:vqa_backdoor_desc}(b). The adversary must estimate or profile these properties to ensure selective activation. Consequently, the backdoor remains dormant under most conditions and is triggered only when the execution environment matches predefined configurations, substantially enhancing stealth.

\begin{figure}[t!]
\centering
\includegraphics[width=2in]{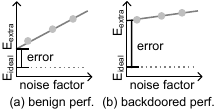}
\caption{The attacking mechanism for ZNE~\cite{Chu:NDSS2026}.}
\label{f:vqa_backdoor_zne}
\vspace{-0.2in}
\end{figure}

\textbf{ZNE manipulation.}
A central component of this attack is the exploitation of \emph{zero-noise extrapolation (ZNE)}~\cite{Chu:NDSS2026}. As exhibited in Figure~\ref{f:vqa_backdoor_zne}(a), ZNE estimates noise-free expectation values by evaluating circuits under scaled noise levels and extrapolating to the zero-noise limit. The attack perturbs circuit parameters so that individual noisy evaluations appear consistent, while the extrapolation process becomes systematically biased, as highlighted in Figure~\ref{f:vqa_backdoor_zne}(b). This results in distorted objective estimates and incorrect optimization outcomes, demonstrating that error mitigation techniques themselves can serve as attack vectors.

\textbf{Robustness and stealth.}
Because the backdoor is encoded in circuit parameters~\cite{Chu:NDSS2026} rather than input triggers or explicit structural modifications, it is robust to compilation, transpilation, and hardware mapping. The model exhibits benign behavior during validation and does not depend on poisoned data, allowing it to evade conventional defenses. Malicious behavior manifests only under specific runtime conditions, e.g., certain qubit regions or noise patterns~\cite{Chu:NDSS2026}, creating a deployment-time vulnerability.

\textbf{Summary.}
This approach builds a hardware-aware, parameter-based backdoor paradigm for VQCs. It removes the need for data poisoning, generalizes to optimization-centric applications, survives compilation, and exploits error mitigation procedures such as ZNE, representing a more realistic and potent threat model in NISQ settings.

\section{Watermarking}

\textbf{Backdoor-based watermarking.}
Prior work~\cite{chu2025bvqc} proposes a backdoor-based watermarking framework, termed BVQC, for intellectual property protection of VQCs. Unlike adversarial backdoor attacks, this work leverages backdoor mechanisms constructively to encode ownership information. The approach embeds a hidden watermark into a pretrained VQC by jointly optimizing the original task objective and a watermark-specific objective through a multi-task learning formulation (see Equation~\ref{e:loss_multi_all}). The watermark is activated only when a set of secret inputs (or quantum states) and corresponding measurements are applied, causing the VQC to produce predefined abnormal outputs that can be used to verify ownership.

\textbf{Mechanism and properties.}
BVQC~\cite{chu2025bvqc} does not rely on data poisoning; instead, it modifies the training objective to incorporate watermark constraints directly into the parameter optimization process. The watermark remains dormant under normal inputs, ensuring that the VQC maintains high performance on its primary task. Importantly, the watermark is designed to be robust to common quantum transformations, including circuit recompilation and transpilation, as it is encoded in the parameter space rather than specific gate structures. The effectiveness of the watermark is evaluated using metrics such as probabilistic proof of authorship (PPA) and ground truth distance (GTD), demonstrating reliable ownership verification with negligible impact on task accuracy.


\section{Defense}

\textbf{Detection scope limitation.}  
Prior work proposes detection frameworks for backdoor attacks in VQC; however, their effectiveness is largely restricted to \emph{classical neural network-like backdoors}~\cite{wang2025qsentry} and certain \emph{compiler-level trojans}~\cite{Das:ISQED2024}. These approaches do not generalize to more advanced, quantum-specific backdoor mechanisms that exploit intrinsic properties of VQCs, such as parameter landscapes, noise characteristics, or error mitigation processes.

\textbf{QSentry (output-based detection).}  
QSentry~\cite{wang2025qsentry} is a defense framework that detects backdoor inputs by analyzing measurement output distributions of quantum neural networks. The method clusters measurement statistics and identifies outliers as potential triggered samples. This approach is effective for \emph{input-trigger-based attacks}, where poisoned inputs induce statistically distinguishable outputs. However, it assumes that backdoor activation produces observable deviations in measurement distributions, an assumption that does not hold for more subtle quantum-specific attacks. In particular, attacks targeting optimization objectives (e.g., VQE, QAOA) or exploiting error mitigation techniques may only manifest under specific execution conditions or during post-processing.

\textbf{TrojanNet (structure-based detection).}  
TrojanNet~\cite{Das:ISQED2024} is a supervised detection framework that classifies quantum circuits (e.g., QAOA instances) as benign or trojaned based on structural representations. Circuit descriptions are encoded into classical feature spaces (e.g., gate sequences or connectivity graphs), and a convolutional neural network is trained to identify anomalous patterns introduced by trojan insertion. This method is effective for detecting \emph{explicit structural modifications}, such as additional gates or altered circuit layouts, which are typical of compiler-level or circuit-insertion attacks. However, it inherently depends on observable structural deviations and is therefore ineffective against backdoors that preserve circuit topology, such as parameter-level manipulations or noise-triggered behaviors.

\textbf{Inability to detect quantum-specific backdoors.}  
Both QSentry~\cite{wang2025qsentry} and TrojanNet~\cite{Das:ISQED2024} rely on classical backdoor signatures, i.e., input-trigger-induced output deviations and structural anomalies. They do not account for \emph{quantum-specific attack vectors}, including parameter-transfer backdoors, noise-triggered activation, or attacks targeting error mitigation (e.g., zero-noise extrapolation). Such backdoors can preserve circuit structure and exhibit benign output distributions under standard conditions, thereby evading both output- and structure-based detection. Consequently, these methods are insufficient for identifying emerging backdoor threats unique to quantum computing.


\vspace{-10pt}
\section{Future Direction}

\textbf{Emerging quantum-native attacks.}
The evolution of backdoor attacks in VQCs is expected to progress toward increasingly \emph{quantum-native} and \emph{environment-aware} mechanisms. Future attacks will likely move beyond input-trigger and structural manipulation parad-igms, instead exploiting deeper properties of quantum systems, including parameter landscapes, noise characteristics, compilation processes, and error mitigation procedures. In particular, adaptive backdoors that respond dynamically to hardware conditions—such as noise levels, calibration drift, qubit connectivity, and native gate sets—may enable highly selective and stealthy activation. Furthermore, as hybrid quantum--classical workflows become more integrated, attackers may target the entire computational stack, from data preprocessing and parameter optimization to compilation and execution. Such multi-stage or cross-layer attack strategies could significantly enhance robustness against transformations and make detection substantially more challenging, especially in heterogeneous and evolving NISQ environments.

\textbf{Toward holistic quantum-aware defenses.}
In response, defense mechanisms must evolve toward \emph{quantum-aware} and \emph{holistic} security frameworks. Future defenses should incorporate multi-level analysis, combining circuit structure inspection, parameter-space auditing, and runtime behavior monitoring under diverse hardware and noise conditions. Techniques such as randomized compilation, cross-device validation, and noise-aware testing may help expose environment-dependent backdoors that evade static analysis. In addition, formal verification and statistical certification methods could provide stronger guarantees on VQC behavior, particularly for safety-critical applications. Importantly, defenses must also consider the interaction between quantum algorithms and error mitigation techniques, as these components can themselves be exploited as attack vectors. Overall, securing VQCs requires a system-level perspective that spans the full quantum software--hardware stack, rather than relying on isolated or single-layer detection approaches.

\vspace{-8pt}
\section{Conclusion}
This paper surveys backdoor attacks in VQCs and categorizes them into data-poisoning, compiler-level, and quantum-native approaches. While early methods resemble classical attacks, recent work exploits quantum-specific properties such as compilation, noise, and parameter transfer. Existing defenses remain limited and largely ineffective against quantum-native threats. Securing VQCs requires quantum-aware, system-level approaches and remains an open research challenge.

\begin{acks}
This work was supported in part by NSF OAC-2417589 and NSF CNS-2143120. We thank the IBM Quantum Researcher \& Educators Program for their support of Quantum Credits.
Any opinions, findings and conclusions or recommendations expressed in this material are those of the authors and do not necessarily reflect the views of grant agencies or their contractors.
\end{acks}

\bibliographystyle{ACM-Reference-Format}
\bibliography{ZNE}
\end{document}